\documentclass[floatfix,lengthcheck,showpacs,amssymb,amsmath,amsfonts,twocolumn]{aastex62}

\usepackage{lineno}
\usepackage[utf8]{inputenc}
\usepackage{color}
\usepackage{graphicx}
\usepackage{float}
\usepackage{subfigure}
\usepackage{amsmath}
\usepackage{acronym}
\usepackage{multirow}
\usepackage{tikz}
\usetikzlibrary{arrows,shapes,trees,decorations.pathreplacing}
\usepackage{import}
\usepackage{svn-multi}
\usepackage{lineno}
\usepackage{hyperref}
\usepackage{gensymb}
\hypersetup{
    colorlinks=true,
    linkcolor=blue,
    filecolor=magenta,      
    urlcolor=cyan,
}

\newcommand{\ecc}{\ensuremath{e}}
\newcommand{\eccthirty}{\ensuremath{e_{30}}}
\newcommand{\chieff}{\ensuremath{\chi_{\mathrm{eff}}}}
\newcommand{\rankingstat}{\ensuremath{\tilde{\rho}_c}}

\newcommand{\msun}{\ensuremath{\mathrm{M}_{\odot}}}

\newcommand{\release}{\texttt{\url{www.github.com/gwastro/eccentric-bns-search}}}

\begin{document}
\title[]{Search for Eccentric Binary Neutron Star Mergers in the first and second observing runs of Advanced LIGO}

\correspondingauthor{Alexander H. Nitz}
\email{alex.nitz@aei.mpg.de}

\author[0000-0002-1850-4587]{Alexander H. Nitz}
\affil{Max-Planck-Institut f{\"u}r Gravitationsphysik (Albert-Einstein-Institut), D-30167 Hannover, Germany}
\affil{Leibniz Universit{\"a}t Hannover, D-30167 Hannover, Germany}

\author[0000-0001-8429-2458]{Amber Lenon}
\affil{Department of Physics and Astronomy, West Virginia University, Morgantown WV 26506, USA}

\author[0000-0002-9180-5765]{Duncan A. Brown}
\affil{Department of Physics, Syracuse University, Syracuse NY 13244, USA}
\affil{Kavli Institute for Theoretical Physics, University of California, Santa Barbara, CA 93106, USA}

\keywords{gravitational waves -- neutron stars -- elliptical orbits}

\begin{abstract}
We present a search for gravitational waves from merging binary neutron stars which have non-negligible eccentricity as they enter the LIGO observing band. We use the public Advanced LIGO data which covers the period from 2015 through 2017 and contains $\sim164$ days of LIGO-Hanford and LIGO-Livingston coincident observing time. The search was conducted using matched-filtering using the PyCBC toolkit. We find no significant binary neutron star candidates beyond GW170817, which has previously been reported by searches for binaries in circular orbits. We place a 90 \% upper limit of $\sim1700$  mergers $\textrm{Gpc}^{-3} \textrm{Yr}^{-1}$ for eccentricities $\lesssim 0.43$ at a dominant-mode gravitational-wave frequency of 10 Hz. The absence of a detection with these data is consistent with theoretical predictions of eccentric binary neutron star merger rates. Using our measured rate we estimate the sensitive volume of future gravitational-wave detectors and compare this to theoretical rate predictions. We find that, in the absence of a prior detection, the rate limits set by six months of Cosmic Explorer observations would constrain all current plausible models of eccentric binary neutron star formation.
\end{abstract}

\section{Introduction}
\label{sec:intro}

\begin{figure*}
  \centering
    \includegraphics[width=2\columnwidth]{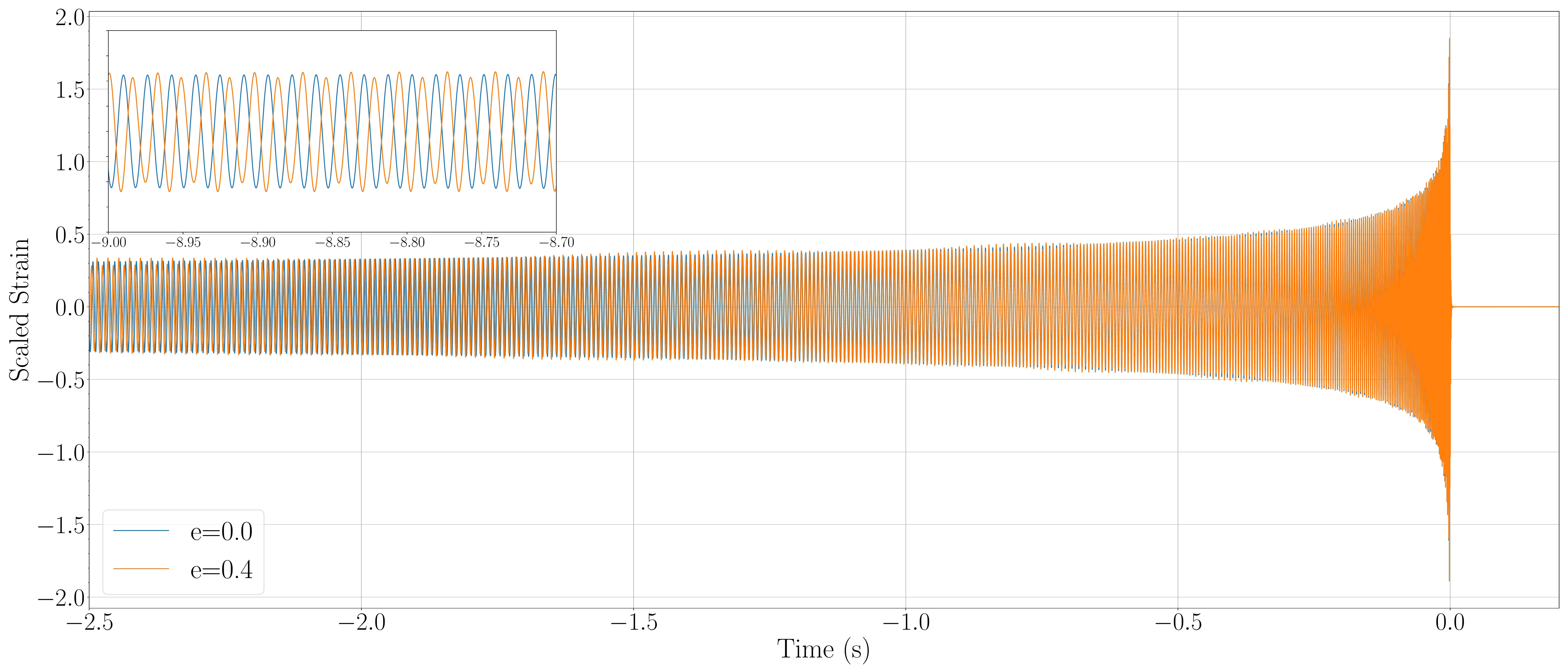}
    \label{fig:waveform}
\caption{EccentricFD gravitational waveforms generated at a dominant-mode gravitational-wave reference frequency of 10Hz with component masses of 1.3$\msun$ for a non-eccentric, e=0.0, (blue) and eccentric, e=0.4, (orange) merging binary neutron star up to the time of merger. Though the waveforms look similar they overlap by $\sim 16\%$. The inset plot shows a zoomed-in depiction of the the phase difference in the non-eccentric (blue) and eccentric (orange) waveforms from -9.0 to -8.7s. 
}
\end{figure*}

With the detections made by the Advanced LIGO (Laser Interferometer Gravitational Wave Observatory)~\citep{TheLIGOScientific:2014jea} and Virgo observatories~\citep{TheVirgo:2014hva}, we have entered the age of gravitational-wave astronomy. During their first (O1) and second (O2) observing runs, the LIGO and Virgo collaborations detected ten binary black hole (BBH) mergers and one binary neutron star (BNS) merger~\citep{LIGOScientific:2018mvr}. Independent groups have since verified these events and detected several additional binary black hole mergers~\citep{Venumadhav:2019tad,Venumadhav:2019lyq,Nitz:2018imz,Nitz:2019hdf}. One possible channel for the formation of merging binaries is through dynamical interaction in dense stellar environments such as globular clusters~\citep{Sigurdsson:1993zrm,PortegiesZwart:1999nm,Grindlay:2005ym} or galactic nuclei~\citep{Oleary:2008myb,Antonini:2012ad}. Unlike binaries formed in the field which can radiate away their eccentricity~\citep{Peters:1964,Hinder:2007qu}, dynamically formed binaries may still have significant residual eccentricity when their gravitational waves enter the LIGO-Virgo band. The observation of a binary with measurable eccentricity would confirm the existence of a dynamical formation channel. The existing LIGO-Virgo BBH candidates are consistent with non-eccentric binary mergers~\citep{Romero-Shaw:2019itr}. The third LIGO-Virgo observing run is currently underway\footnote{https://gracedb.ligo.org/superevents/public/O3} and is expected to produce dozens more events~\citep{Aasi:2013wya}. 

A search for eccentric BBH mergers in O1 and O2 data using methods which do not use models of the gravitational waveform~\citep{Klimenko:2008fu,Klimenko:2015ypf,Tiwari:2015gal} reported no eccentric merger candidates~\citep{Salemi:2019owp}. The sensitivity of gravitational-wave searches can be improved by the use of matched-filtering, if a model of the target waveform is available. Existing matched-filter searches were designed for the detection of circular  binaries~\citep{DalCanton:2017ala,Usman:2015kfa,Venumadhav:2019tad}. It is possible that compact binaries with measurable eccentricity may have been missed by these initial searches~\citep{Brown:2009ng,Huerta:2013qb}. For BBH mergers, highly accurate models with the full inspiral-merger-ringdown, along with support for both a large range of eccentricity and spin do not yet exist, though development is rapidly progressing and there are models which satisfy some of these constraints~\citep{Huerta:2017kez,Cao:2017ndf,Hinderer:2017jcs,Hinder:2017sxy,Ireland:2019tao}.

In this paper, we search for eccentric BNS mergers.  There are several models of the gravitational waveform suitable for this task which include EccentricFD~\citep{Huerta:2014eca} and TaylorF2e~\citep{Moore:2018kvz,Moore:2019xkm}. These waveform models do not currently support compact-object spin. However, neutron star binaries formed by dynamical capture in globular clusters may have non-negligible spin if they follow the observed distribution of millisecond pulsars (MSPs). Even if large spins are supported, we may expect the  effective spin $\chieff = (\chi_{1z} m_1 + \chi_{2z} m_2)/(m_1+m_2)$ to peak around zero if the individual neutron stars orientations are isotropic. Searches which do not account for spin still have significant sensitivity to sources with low effective spin $\chieff < 0.1$, though there will be significantly reduced sensitivity in the case where both component neutron stars are consistent with the fastest observed MSP~\citep{Hessels:2006ze} and their respective spins are aligned with the orbital angular momentum~\citep{Brown:2012gs}.

Using these waveforms, we perform a matched-filtering based analysis by extending the methods used by \cite{Nitz:2018imz} to include eccentric binaries. We find that our search is effective at detecting eccentric BNS mergers up to an eccentricity \ecc~$\sim 0.43$ at a dominant-mode gravitational-wave frequency of 10 Hz. Using a representative sample of the O1 and O2 dataset, we find that a non-eccentric search starts losing significant sensitivity relative to the eccentric search starting at $\ecc \sim 0.07$, in agreement with the results of~\cite{Huerta:2013qb,Moore:2019vjj}.

We find no individually significant eccentric BNS merger candidates using the public O1 and O2 datasets~\citep{Vallisneri:2014vxa}. The only significant event is the previously reported merger GW170817~\citep{TheLIGOScientific:2017qsa} since our search is also sensitive to circular binaries. In the absence of a new detection, we place a $90\%$ upper limit on the merger rate of $\sim 1700~\textrm{Gpc}^{-3}\textrm{Yr}^{-1}$ for binaries whose eccentricity is \ecc~$\lesssim0.43$ at the 10 Hz reference frequency. While we do not detect any individually significant mergers, it is possible that follow-up could uncover sub-threshold sources, and so we make available our full population of sub-threshold candidates~\citep{1-ECCBNS}.

We can compare our measured rate to predictions for the proposed channels for eccentric BNS formation. \cite{Lee:2009ca} predict a BNS merger rate of 30 $\textrm{Gpc}^{-3} \textrm{Yr}^{-1}$ at z=0 from binaries formed by the tidal capture and collision of neutron stars in globular clusters. \cite{Ye:2019xvf} predict a merger rate of $\sim 0.02 \textrm{Gpc}^{-3} \textrm{Yr}^{-1}$ from binaries formed by dynamical interactions in globular clusters. Given these predicted rates, it is unsurprising that our search did not observe a signal. Future detectors like A+~\citep{Aasi:2013wya} and Cosmic Explorer~\citep{Reitze:2019iox}, will observe a large volume of the universe and have a higher probability of observing eccentric BNS mergers. Using our measured rate and the expected sensitivity of A+ and Cosmic Explorer, we estimate the time it would take for observed rates to impinge on the predicted rates. Using the A+ expected sensitivity distance of 330 Mpc~\citep{Aasi:2013wya}, we find the most optimistic predictions~\citep{Lee:2009ca} require half a year of data for the measurement to be comparable to the predictions. The most pessimistic predictions~\citep{Ye:2019xvf}  require $\sim 775$ years of data before the measured rate limits are comparable with the prediction. However, the proposed third-generation detector Cosmic Explorer would need at most half a year of data to achieve a rate limit comparable to the most pessimistic models, although a serendipitous detection is always a possibility with  current detectors.

\section{Search Methodology}   

We use a matched-filtering search for compact-object binaries  using the PyCBC toolkit~\citep{pycbc-github}. We use gravitational waveforms that model mergers with elliptical orbits, but otherwise employ the same configuration as used by~\cite{Nitz:2018imz} for their search for gravitational waves from compact binary mergers.

\begin{figure}[t]
  \centering
    \includegraphics[width=1.2\columnwidth]{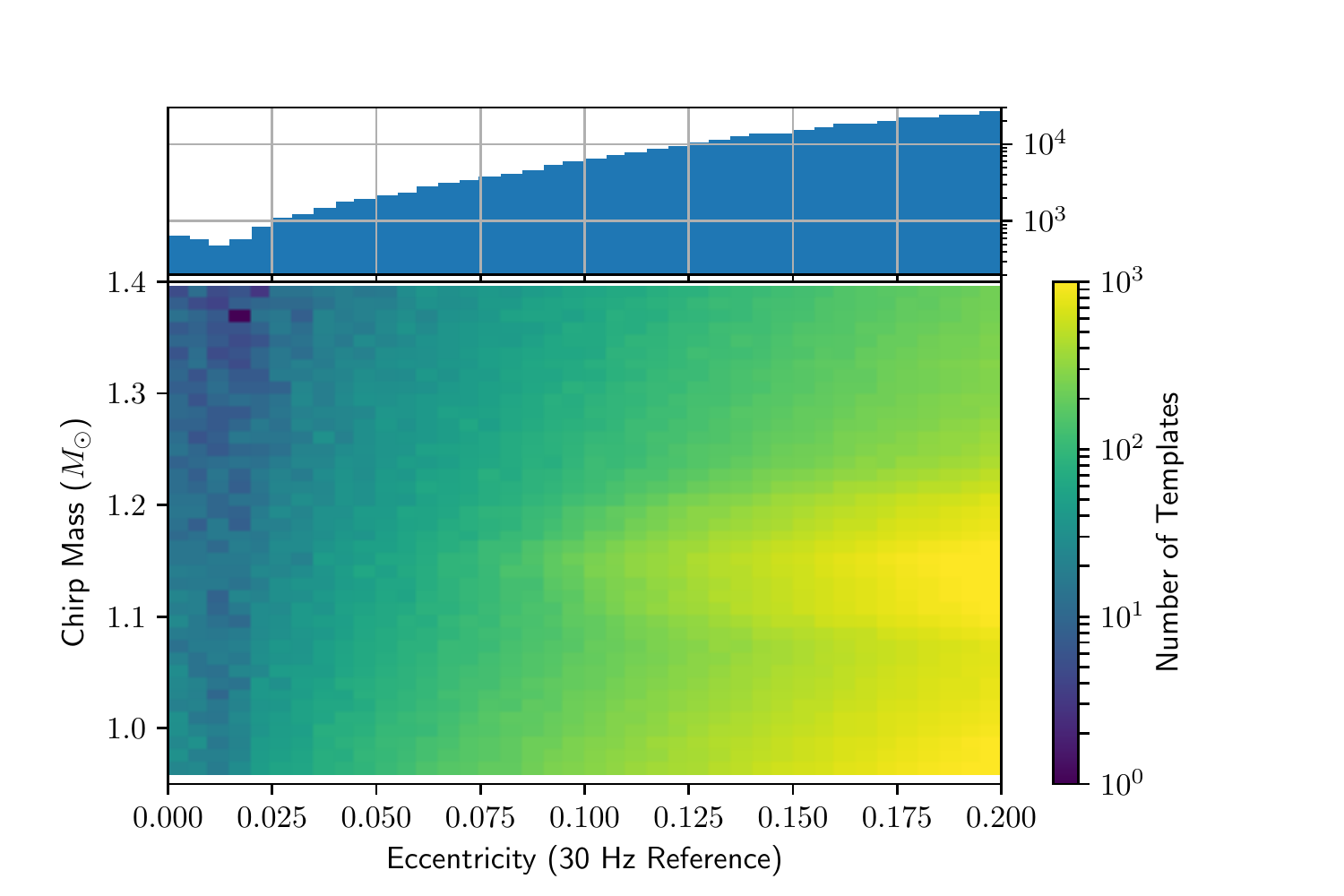}
    \label{fig:bank}
\caption{This distribution of templates in our eccentric BNS bank. Note that the eccentricity
is given at a dominant-mode gravitational-wave reference frequency of 30 Hz as opposed to 10 Hz used elsewhere in this paper.
}
\end{figure}

Of the available waveform models, we employ two waveform models that contain eccentricity, EccentricFD and TaylorF2e. EccentricFD~\citep{Huerta:2014eca} extends the post-circular (PC) analysis of \cite{Yunes:2009yz} to obtain a 3.5PN Fourier-domain enhanced PC gravitational-wave model that produces an eccentric, compact binary inspiral waveform in the small eccentricity approximation.  In the zero eccentricity limit this model reproduces the non-eccentric model, TaylorF2, and in the small eccentricity limit this model will reproduce the PC model to leading order. Fig.~\ref{fig:waveform} shows two waveforms generated using EccentricFD with a non-eccentric waveform shown in blue and an eccentric waveform shown in orange. TaylorF2e is a 3PN Fourier-domain, eccentric waveform model, valid for larger initial eccentricities, defined by the stationary phase approximation (SPA) of a harmonically-decomposed time-domain signal. While both models expand the amplitude coefficients in small eccentricity, the TaylorF2e model does not invert the dependence of orbital frequency on eccentricity and numerically solves the stationary phase condition~\citep{Moore:2018kvz,Moore:2019xkm,Moore:2019vjj}.

We find that a template bank generated by straightforward stochastic placement of EccentricFD waveforms starting at a gravitational-wave frequency of 30 Hz is sufficient to recover BNS signals with eccentricity as modelled by either EccentricFD or TaylorF2e. In addition to the component masses of the BNS, our bank adds a parameter for the eccentricity, $\eccthirty$, along with an additional binary orientation parameter. Our template bank is designed to detect BNS mergers where the component masses range from $1.1-1.6 \msun$~\citep{Ozel:2016oaf} and eccentricities up to 0.2 at a reference of 30 Hz. This corresponds to $\sim0.43$ at a reference frequency of 10 Hz. Fig.~\ref{fig:bank} shows the distribution of templates in both chirp mass and eccentricity. The density of templates increases rapidly with eccentricity. Adding the additional degrees of freedom increases the size of the template bank by a factor of $160$ relative to a non-eccentric, non-spinning bank that would cover the same region. Due to the inherent degeneracy between the component masses, the template bank will have significant sensitivity outside this parameter space in regions where the chirp mass $\mathcal{M} = (m_1m_2)^{3/5} / (m_1+m_2)^{1/5}$ is otherwise consistent i.e. a $1.2 - 2.0\msun$ merger. 

Matched-filtering is used to calculate the signal-to-noise (SNR) time series using our bank of template waveforms independently for each observatory~\citep{Allen:2005fk}. Peaks in the SNR time series are followed up by a series of signal consistency tests~\citep{Nitz:2017lco,Allen:2004gu} and combined into multi-detector candidates~\citep{Usman:2015kfa,Nitz:2017svb}. We assign each candidate a ranking statistic, \rankingstat, using the same methods employed in the 1-OGC catalog~\citep{Nitz:2018imz}. The ranking statistic, \rankingstat, accounts for the signal-to-noise (SNR) of each candidate, the consistency of its morphology and signal properties with an astrophysical source, and the rate of background for candidates arising from similar templates.

\section{Observational Results}
\begin{table*}
  \begin{center}
    \caption{Binary neutron star candidates from the search of O1 and O2 LIGO data sorted by the rate of false alarms with a detection statistic at least as large as the candidate. The mass and eccentricity parameters of the template associated with each candidate are listed. Note the eccentricity is given at the 30 Hz gravitational-wave frequency reference used to generate the template bank. The values associated with a candidate can be considered point estimates and may differ significantly from the results of full Bayesian parameter estimation. Masses are quoted in the detector frame.}
    \label{table:search}
\begin{tabular}{ccrcrrrrrrrr}
Date designation & GPS time & FAR$^{-1}$ (y) & $\rankingstat$ & $\rho_H$ & $\rho_L$ & $m_1$ & $m_2$ & $\eccthirty$\\ \hline
170817+12:41:04UTC          & 1187008882.45          &   $>10000$                     &      27.86          &      18.41          &      23.60          &       1.48          &       1.28          &       0.02         \\
151127+02:24:56UTC          & 1132626313.67          &        .57                     &       8.60          &       7.28          &       5.73          &       1.23          &       1.55          &       0.16         \\
151130+22:40:53UTC          & 1132958470.76          &        .54                     &       8.60          &       6.76          &       5.89          &       1.29          &       1.22          &       0.19         \\
170705+12:02:50UTC          & 1183291388.00          &        .31                     &       8.54          &       7.29          &       5.56          &       1.48          &       1.57          &       0.16         \\
151227+13:12:35UTC          & 1135257172.28          &        .14                     &       8.42          &       6.33          &       6.21          &       1.42          &       1.37          &       0.10         \\
170618+15:35:01UTC          & 1181835319.00          &        .08                     &       8.40          &       7.30          &       5.35          &       1.22          &       1.19          &       0.15         \\
170812+20:07:43UTC          & 1186603681.67          &        .07                     &       8.35          &       6.92          &       5.47          &       1.21          &       1.13          &       0.17         \\
170302+22:45:10UTC          & 1172529928.62          &        .07                     &       8.42          &       6.93          &       5.46          &       1.23          &       1.17          &       0.12         \\
161222+07:49:11UTC          & 1166428168.98          &        .06                     &       8.39          &       6.33          &       6.14          &       1.50          &       1.12          &       0.18         \\
170328+07:26:40UTC          & 1174721218.74          &        .05                     &       8.38          &       5.11          &       7.26          &       1.11          &       1.22          &       0.12         \\
\end{tabular}
  \end{center}
\end{table*}

We search the public LIGO O1 and O2 dataset which contains $\sim164$ days of coincident LIGO-Hanford and LIGO-Livingston data after removal of data which has been flagged as potentially containing instrumental artefacts~\citep{TheLIGOScientific:2016zmo,TheLIGOScientific:2017lwt,Vallisneri:2014vxa}. Data when only a single observatory was operating was not considered, nor was data from the Virgo observatory which operated only in the last month of O2. In this search, we neglect data from the Virgo detector as it only provides a marginal senstivity improvement~\citep{Nitz:2019hdf}. Future analyses will incorporate data from the full network.

The most significant candidates are listed in Table~\ref{table:search}. As our search is also sensitive to circular binaries, it is not surprising that GW170817---first detected by the LIGO-Virgo search for circular binaries---was observed as a high-significance event. The remaining candidates are consistent with the rate of false alarms expected for the amount of data analyzed. However, we cannot rule out a sub-threshold population which may be uncovered by correlation with non-GW datasets (GRBs, Kilonovae, etc) such as performed in~\cite{Nitz:2019bxt}.

\section{Upper Limits}

\begin{figure}[t]
  \centering
    \includegraphics[width=1.1\columnwidth]{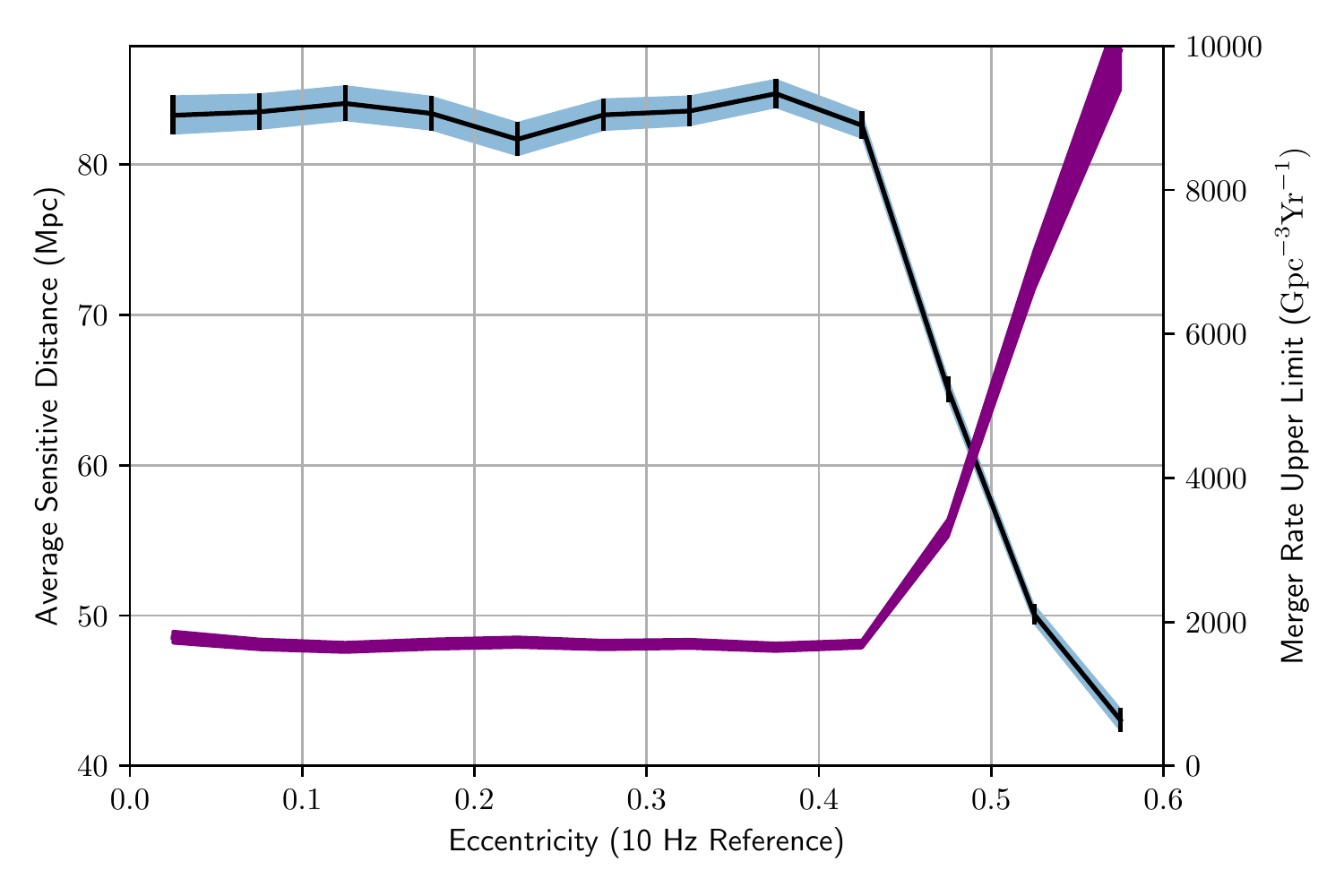}
\caption{The average sensitive distance of the search (blue/left scale) and the 90$\%$ upper limit on the rate of eccentric binary neutron star mergers (purple/right scale) as a function of eccentricity at a reference frequency of 10 Hz. The average sensitivity is nearly flat up to an eccentricity of 0.43, where we begin to see sharp drop-off in sensitive range. This corresponds to the edge of our template bank.
}
\label{fig:range}
\end{figure}

As our search did detect any significant individual eccentric BNS merger candidates, we place an upper limit on the rate of eccentric mergers as a function of their eccentricity. We determined a $90\%$ confidence upper limit on the rate of mergers using the method introduced in~\cite{Brady:2004gt}. The upper limit on the merger rate $R_{90}$ is

\begin{equation}
    R_{90} = 2.303 \left[TV(\mathcal{F}^*)\right]^{-1}
\end{equation}

where $T$ is the total observation time and $V(\mathcal{F^*})$ is the average volume the search is sensitive to at the false alarm rate of the loudest observed candidate. Under the assumption that GW170817 is a non-eccentric merger, we exclude it from our analysis. The sensitivity is measured using a simulated population of sources distributed uniform in volume and isotropic in orientation. We have primarily used the EccentricFD model for our simulated population, however, we have confirmed our results are consistent with a smaller sample using the TaylorF2e model. Fig.~\ref{fig:range} shows the upper limit on the merger rate as a function of the binary eccentricity as well as the average sensitive distance of the search over the observation period. We find that up to an eccentricity of $\sim 0.43$ at a reference frequency of 10 Hz, we can place a $90\%$ upper limit at $\sim$1700 mergers per cubic Gpc per year.

Under the assumption that eccentric signals will not have been detected, we can determine the observation time required by future detectors to constrain the BNS merger rates predicted by~\cite{Lee:2009ca} and \cite{Ye:2019xvf} by scaling the upper limit from our search. We find that the Advanced LIGO observatories had an average range, $D_{O1+O2}$, of 90 Mpc during O1 and O2 for a fiducial $1.4-1.4 \msun$ merger by taking the weighted-average of their noise curves. Similarly, using their respective noise curves, we find an average range, $D_{A+}$, of 330 Mpc for A+~\citep{Aasi:2013wya} and $D_{CE}$ of 7130 Mpc for Cosmic Explorer\footnote{https://cosmicexplorer.org/researchers.html}. The observation time required, $T_{CE,A+}$, to match the predicted rates, $R_{Ye, Lee}$, is given as
\begin{equation}
    T_{CE,A+|Ye,Lee} = T_{O1+O2} \frac{R_{O1+O2}}{R_{Ye,Lee}} \left(\frac{D_{O1+O2}}{ D_{CE,A+}}\right)^3,
\end{equation}
where $T_{O1+O2}$ is the total observation time of O1 and O2 and $R_{O1+O2}$ is the upper limit achieved by our current search. We find that with the increased sensitivity of A+ the most optimistic predictions~\citep{Lee:2009ca} would require half a year of data and the most pessimistic predictions~\citep{Ye:2019xvf} would require $\sim 775$ years. Cosmic Explorer would need at most half a year of data to constrain current BNS merger rate models. Understanding the constraints that future observational limits place on eccentric binary formation channels will require computation of the rate as a function of eccentricity from population synthesis.

\section{Conclusions}

We have developed a search that is effective at detecting BNS mergers with orbital eccentricity $\lesssim0.43$ at 10 Hz. Our search uses the public PyCBC toolkit~\citep{pycbc-github} based on a standard matched filtering approach~\citep{Nitz:2018imz,Usman:2015kfa}. We have found that straightforward stochastic placement algorithms are sufficient to tackle the construction of template banks for eccentric binary merger waveforms. As broadly applicable and highly accurate eccentric waveform models are developed which include corrections for component-object spin, the full inspiral-merger-ringdown, and support for large values of eccentricity it will be possible to apply the same methods demonstrated here to the detection of BBH mergers.

To aid in further analysis of our results, we make available our full sub-threshold catalog of eccentric BNS candidates. For each candidate we provide the false alarm rate, parameters of the associated template waveform, and signal parameters such as the signal-to-noise and results of our signal consistency tests~\citep{1-ECCBNS}\footnote{\release}.

While the detection of a single BNS or BBH eccentric merger would immediately demonstrate the existence of dynamical formation, current estimates of the rate of BNS mergers imply that a single observation would be rare for the current generation of ground based observatories. Future observatories such as Cosmic Explorer will be able to probe current models.

\label{sec:disc}

\acknowledgments
  We thank Nico Yunes and Blake Moore for their feedback and guidance using TaylorF2e. DAB thanks National Science Foundation Grant No.~PHY-1707954 for support. AL thanks The Center for Gravitational Waves and Cosmology at West Virginia University, and the Division of Diversity Equity and Inclusion at West Virginia University for support. We acknowledge the Max Planck Gesellschaft for support and the Atlas cluster computing team at AEI Hannover. This research was supported in part by the National Science Foundation under Grant No.~PHY-1748958. This research has made use of data, software and/or web tools obtained from the Gravitational Wave Open Science Center (https://www.gw-openscience.org), a service of LIGO Laboratory, the LIGO Scientific Collaboration and the Virgo Collaboration. LIGO is funded by the U.S. National Science Foundation. Virgo is funded by the French Centre National de Recherche Scientifique (CNRS), the Italian Istituto Nazionale della Fisica Nucleare (INFN) and the Dutch Nikhef, with contributions by Polish and Hungarian institutes.\newline\newline\newline
\bibliography{references}

\end{document}